\documentclass[10pt,conference]{IEEEtran}

\usepackage[T1]{fontenc}
\usepackage[utf8]{inputenc}
\usepackage{microtype}
\usepackage{booktabs}
\usepackage{tabularx}
\usepackage{array}
\usepackage[dvipsnames]{xcolor}
\usepackage{xspace}
\usepackage{enumitem}
\usepackage{xcolor}
\usepackage{tikz}
\usetikzlibrary{positioning,arrows.meta}
\usepackage[hidelinks]{hyperref}
\usepackage{todonotes}
\usepackage{graphicx}
\usepackage{subcaption}
\usepackage{comment}

\usepackage{xcolor}
\newcommand{\dataset}{\textsc{Story2Diff}\xspace}
\newcommand{\dataseturl}{https://github.com/riganelli/user-story-mining-pipeline/releases/tag/V4.0.1}
\newcommand{\codeurl}{https://github.com/riganelli/user-story-mining-pipeline}
\newcommand{\videourl}{https://youtu.be/qOhXUeyeiTQ}

\newcommand{\numStories}{7,785\xspace}
\newcommand{\numRepositories}{982\xspace} 

\title{\dataset: A Dataset Linking Real-World User Stories to Code Changes}

\author{
\IEEEauthorblockN{Benedetta Donato, Daniele Franchini, Alessandro Greco, Oliviero Riganelli, Daniela Micucci, Leonardo Mariani}
\IEEEauthorblockA{
University of Milano-Bicocca, Milan, Italy\\
\{benedetta.donato,oliviero.riganelli,daniela.micucci,leonardo.mariani\}@unimib.it, \{d.franchini2,a.greco71\}@campus.unimib.it
}}

\begin{document}
\maketitle

\begin{abstract}
Research on requirements-to-code tasks needs examples in which a genuine requirement is paired with the change that implemented it. Existing user-story collections, however, largely contain isolated or synthetic text. To address this limitation, we present \dataset, a dataset of \numStories English user stories linked to merged pull requests from January 2017 to July 2026. The dataset is mined from a set of \numRepositories public Java and Python repositories. Each record in the dataset combines the original issue, extracted Connextra elements, the associated pull request and commit metadata, a file-level diff capturing the code changes introduced by the associated pull request,
and the affected files before and after the change. The tool we built to mine \dataset can be used to repeat the automatic extraction of the records, using different parameters (e.g., a different timeframe).
The dataset can support a variety %
of tasks, including requirement discovery, traceability, change-impact estimation, and requirements-aware code generation.
Dataset: \url{\dataseturl}. Video: \url{\videourl}.
\end{abstract}

\begin{IEEEkeywords}
User Stories, Requirements Traceability, Mining Software Repositories
\end{IEEEkeywords}

\section{Introduction}
User stories are commonly used to specify requirements from a stakeholder's perspective, often through the Connextra template: \emph{As a [role], I want [feature], so that [benefit]}~\cite{cohn2004}. They are widely used in agile development processes, such as Scrum and XP.

A wide spectrum of techniques needs both large sets of requirements and their implementation to be properly assessed. This includes automated traceability recovery, which identifies links between requirements and implementation artifacts~\cite{rath2018}; requirements-driven change-impact analysis, which identifies the code affected by a requirement change~\cite{Falessi2020}; and code generation and modification from natural-language requirements~\cite{jimenez2024swebench}. Paired requirements and implementations provide reference artifacts against which recovered links, predicted changes, and generated code can be assessed.

Existing user-story datasets generally contain collections of stories without associated implementations
~\cite{lucassen2016,quattrocchi2026,yamani2025}.
Such datasets are sufficient to study, for instance, the linguistic quality of the requirements~\cite{lucassen2016}, but they are \emph{insufficient for investigating questions that cross the boundary between requirements and implementation}. %

To address this gap, we built \dataset, a dataset of user stories linked to the code changes introduced by the pull requests that close their corresponding GitHub issues, together with tool support to repeat and adapt the extraction process to future needs.
In particular, we extract issues that contain user stories matching the Connextra template that are bound to merged pull requests (PRs).

In a nutshell, \dataset delivers \numStories real-world user-story--implementation records spanning \numRepositories Java and Python repositories and nearly ten years of development; and a documented and reproducible multi-stage mining process that combines LLM validation, issue analysis, and text verification.

\section{Related Datasets and Studies}
\noindent\textbf{User-story datasets.} Several corpora collect natural-language user stories, but none pairs
each story with the code change that implements it.
Lucassen~et~al. operationalize syntactic, pragmatic, and semantic
quality criteria on 1{,}023 industrial user stories through the Quality
User Story (QUS) framework~\cite{lucassen2016}.
Quattrocchi~et~al. assess whether LLMs can generate or evaluate user
stories, producing an annotated benchmark for requirement-quality
research~\cite{quattrocchi2026}.
Yamani~et~al. contribute UStAI, a dataset of 1{,}260 LLM-generated
user stories for AI-intensive systems, supporting quality-assessment
research but offering no developer-authored stories or paired code
changes~\cite{yamani2025}.
These resources enable linguistic-quality or generation studies, but
they cannot support empirical work that crosses the boundary between
requirements and implementation.

\noindent\textbf{Traceability datasets.}
Existing traceability datasets link issue or requirement artifacts to
code but rarely adopt user-story framing. Rath~et~al. train a classifier that recovers missing issue-to-commit
links~\cite{rath2018}; SEOSS provides 33~projects with typed
issue-to-issue and issue-to-commit trace links~\cite{rathmader2019}.
PI-Link contains 50,369 explicit issue-to-pull-request links mined
from Android repositories~\cite{alshara2023}, and
Majidzadeh~et~al.\ release issue-to-commit and issue-to-method links
mined from GitHub~\cite{majidzadeh2022}.
Elamin and Osman model user stories, tasks, and developer tests as
typed nodes in a Neo4j graph~\cite{elamin2018qrs}, but their links
never reach code-level artifacts such as commits or diffs.
P\^{a}rtachi~et~al.\ suggest missing PR--issue links online, when a
developer submits a PR or closes an issue~\cite{partachi2023}.
None of these resources anchors the requirement side to canonical Connextra user stories paired with code changes associated with their implementation. %

\noindent\textbf{Code-generation benchmarks.} SWE-bench~\cite{jimenez2024swebench} pairs 2,294 merged PRs from 12~Python
repositories with the issues they resolve, and scores generated patches
by running the project's tests; its requirement is the issue text as
written, typically a bug report or feature request. Subsequent studies report
 leakage of the solution, weak tests~\cite{aleithan2026}, and possible
memorization~\cite{liang2026}. PR4Code~\cite{donato2026pr4code}
collects 13,339 Java and Python PRs with commit sequences and
cumulative diffs, but selects them by keywords, so the textual side is
a free-form PR description, and only 3,603 PRs use issue-closing
syntax. Unlike both, each \dataset record links a user story to the merged PR that closed its issue, and keeps the full commit sequence, file-level diffs, and files before and after the change. Similar to PR4Code, the data collection process can be repeated on a more recent time window, enabling the construction of fresh benchmark instances and thereby reducing the risk of contamination.

\dataset complements prior work by providing a curated collection of user stories extracted from GitHub issues together with their corresponding code changes, materialized as pull requests. This pairing of natural‑language requirements and concrete implementations enables empirical studies that directly examine how user stories relate to code.

\section{Dataset Construction}
\begin{figure*}[t]
\centering
\begin{tikzpicture}[
  node distance=4mm,
  every node/.style={align=center},
  box/.style={
    draw,
    rounded corners,
    minimum height=14mm,
    text width=25mm,
    inner sep=5pt,
    fill=black!3,
    font=\scriptsize
  },
  arrow/.style={-{Latex[length=2mm]},thick}
]

\node[box] (repos) {
  \textbf{Repository Discovery}\\[1mm]
  73,680 examined\\
  5,638 retained
};

\node[box,right=of repos] (issues) {
  \textbf{Issue Extraction}\\[1mm]
  99,194 candidate issues
};

\node[box,right=of issues] (prs) {
  \textbf{Merged-PR Linkage}\\[1mm]
  23,527 linked issues
};

\node[box,right=of prs] (valid) {
  \textbf{Grounded LLM Validation}\\[1mm]
  13,192 validated issues
};

\node[box,right=of valid] (records) {
  \textbf{Code Materialization}\\[1mm]
  13,192 materialized records
};

\node[box,right=of records] (literal) {
  \textbf{Literal Verification}\\[1mm]
  7,785 final records
};

\draw[arrow] (repos) -- (issues);
\draw[arrow] (issues) -- (prs);
\draw[arrow] (prs) -- (valid);
\draw[arrow] (valid) -- (records);
\draw[arrow] (records) -- (literal);

\end{tikzpicture}
\caption{\dataset construction process}
\label{fig:pipeline}
\vspace{-6mm}
\end{figure*}

Figure~\ref{fig:pipeline} summarizes the six-stage process we defined to construct the \dataset dataset. The pipeline first identifies eligible repositories (Repository Discovery), then extracts the candidate issues (Issue Extraction) and associates them with the merged
pull requests that closed them (Merged-PR Linkage). It then uses an LLM to extract user stories from issues (Grounded LLM Validation). Next, it retrieves the implementation artifacts (Code Materialization), and finally, it retains only records whose extracted components form a user story that occurs verbatim in the original issue text (Literal Verification). %

\subsection{Repository Discovery}
GitHub does not provide an explicit repository-level marker for user-story adoption. We therefore identify candidate repositories through five complementary strategies, implemented in Python using the \texttt{requests} library~\cite{requests_docs} and GitHub's REST Search API. %
We apply the strategies sequentially and maintain a shared set of previously encountered repositories to avoid duplicate processing across strategies. Repositories identified by any of the following strategies are considered for subsequent filtering:
\smallskip

\textbf{LABEL}: select repositories with closed issues whose labels indicate the use of user stories by matching \texttt{user-story}, \texttt{user story}, or \texttt{user\_story}. %
    
\textbf{TEMPLATE}: searches for repositories that store issue templates under \texttt{.github/ISSUE\_TEMPLATE} with content that includes ``user story'' or a Connextra statement.
    
\textbf{TAG}: searches closed issue titles with the keyword \texttt{US-}, then applies a case-insensitive regular expression to the returned titles. The expression recognizes four marker families: (i) \texttt{[US]}; (ii) \texttt{[US}$n$\texttt{]}, allowing zero or more hyphens, underscores, or whitespace characters between \texttt{US} and $n$; (iii) \texttt{[userstory]}, allowing zero or more hyphens, underscores, or whitespace characters between \texttt{user} and \texttt{story}; and (iv) \texttt{US}$n$ at the beginning of the title, allowing whitespace and an optional single hyphen or underscore between \texttt{US} and $n$, with a word boundary after $n$. Here, $n$ denotes one or more digits. Surrounding whitespace is permitted inside brackets and before title-initial markers.
    
\textbf{MENTION}: searches for repositories containing issues whose titles or bodies include ``user story'', without requiring a specific sentence structure.

\textbf{CONNEXTRA}: searches for repositories containing issues whose bodies include the definition of a role and a desired action, following the formulation ``As a [role], I want [action]''. Since keyword searches can return unrelated matches, we use client-side regular expressions to refine the results. The expressions recognize both partial stories, containing a role and an action introduced by ``I want'', ``I need'', or ``I would like'', and complete stories that additionally express a benefit through ``so that'', ``in order to'', or ``so I can''. %

\smallskip

For the construction of the dataset, we restrict the scope to Java and Python repositories, two widely used programming languages on GitHub~\cite{GitHut2.0}. We require at least one of these languages to individually account for at least 30\% of the repository's code. Moreover, a repository is retained only if it has at least two stars. This phase examined 73,680 unique repositories and retained 5,638 for subsequent extraction.

\subsection{Issue Extraction}
This stage takes as input the repositories retained during Repository Discovery and identifies individual issues that may contain user stories. %
For each repository, the pipeline runs repository-scoped queries through GitHub's REST Search API (\texttt{/search/issues}), applying the five described discovery strategies and selecting \emph{issues closed} between 1~January 2017 and 4~July 2026 (the mining date). %

For each issue, the pipeline checks the language using the Python \texttt{langid} library. Before language identification, code blocks, URLs, and HTML tags are removed from the issue text to reduce interference from non-natural-language content. Issues classified as non-English with a confidence of at least 0.95 are discarded. %
Issues whose cleaned text contains fewer than 20 characters bypass language identification and are also retained for subsequent processing.

This stage returned a total of 99,194 candidate issues.%

\subsection{Merged-PR Linkage}
Since the objective of the \dataset dataset is to associate requirements written as user stories with their implementations, this stage associates candidate issues with the corresponding merged pull requests. In particular, for each issue, the pipeline uses its repository and issue number to query GitHub's GraphQL API and %
identify the PR responsible for closing the issue, as recorded in the ClosedEvent.closer field.
Only issues closed by a merged PR are retained, excluding those closed by commits or associated with PRs through references that did not result in issue closure. This ensures that the linked changes correspond to code integrated into the repository and provides an explicit traceability link between the user story and the associated code changes. %
Overall, this stage retained 23,527 issues associated with a merged PR that closed them.

\subsection{Grounded LLM Validation} Since the identification of user stories based on regular expressions might be imprecise, we further analyze issue descriptions using LLMs to (a) confirm whether an actual user story is present and (b) extract its individual components. To this end, we run a Qwen2.5 7B  model locally at temperature zero to extract from the issues a JSON object containing: (a) a Boolean verdict (\texttt{is\_user\_story}), indicating whether the issue contains a user story; (b) a classification of the Connextra formulation as \texttt{complete} (role and feature with an explicit benefit), \texttt{partial} (role and feature without an explicit benefit), or \texttt{absent} (otherwise); (c) the extracted role, feature, and benefit, when present; and (d) the \texttt{location} of the user story within the issue, which can be the \texttt{title}, \texttt{body}, or \texttt{both}, with \texttt{null} allowed when no location is reported. %
We retain only issues for which the LLM returns \texttt{is\_user\_story} = true, without applying additional grounding or confidence-based constraints. The literal correspondence between the extracted components and the original issue text is verified in the subsequent Literal Verification stage. Overall, this stage retained 13,192 issues.

\subsection{Code Materialization}
\begin{figure}[ht]
    \centering
    \includegraphics[width=\linewidth]{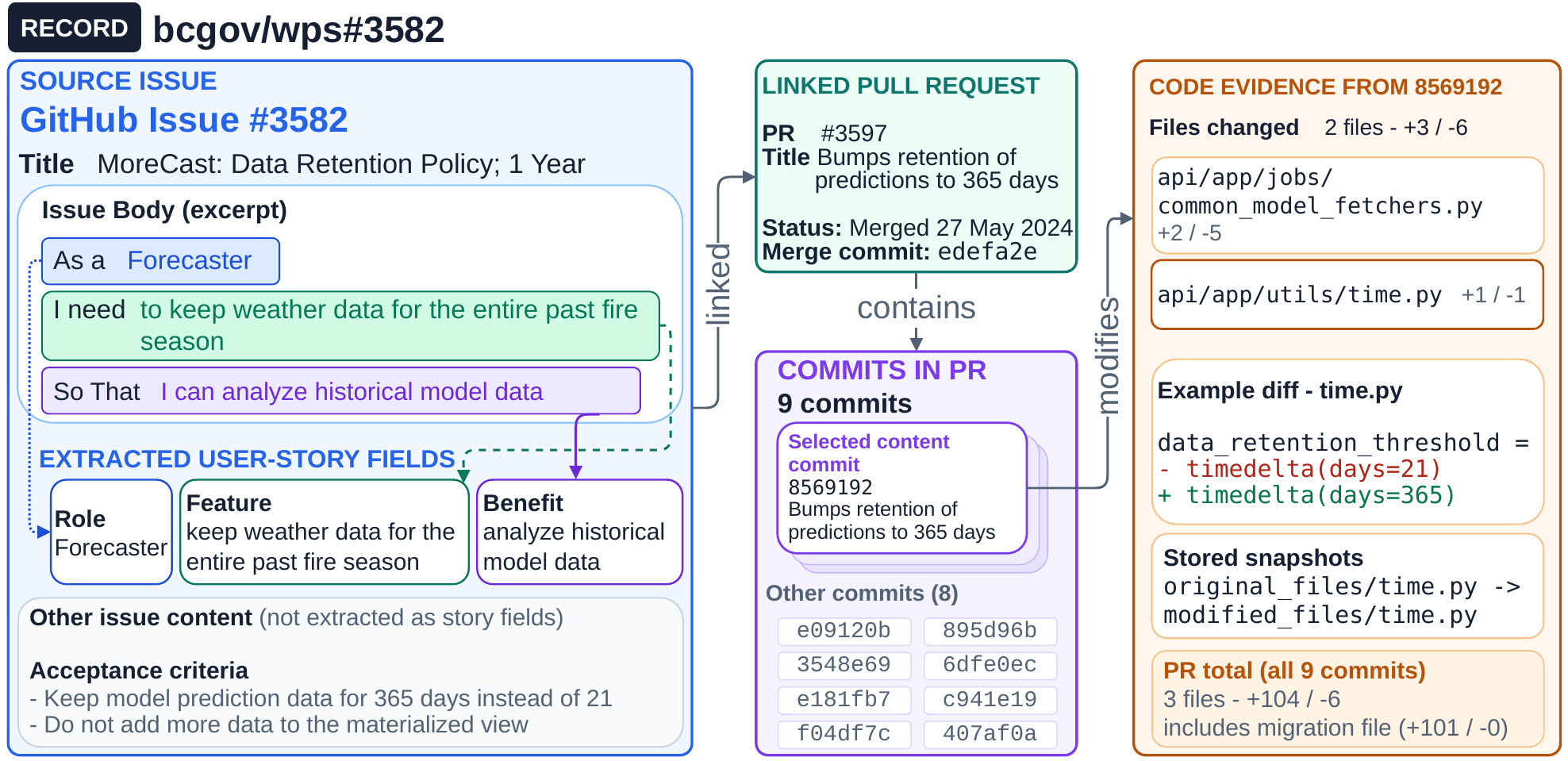}
    \caption{Example record linking the user story in
\texttt{bcgov/wps\#3582} to PR~\#3597, commits,
and changes.}
\label{fig:Story2Diff_record_hierarchical.drawio}
\vspace{-6mm}
\end{figure}

For each issue, this stage retrieves the associated implementation artifacts through GitHub's REST API. Figure~\ref{fig:Story2Diff_record_hierarchical.drawio} visually exemplifies the main data retrieved for each issue and PR, considering the case of issue \texttt{bcgov/wps\#3582} and its PR~\#3597.
The retrieved PR metadata includes its identifier, title, URL, source and target branches, merge timestamp, and the SHA identifiers of the base revision, head revision, and merge commit. The pipeline also retrieves the commit sequence between the base and head revisions (i.e., the full commit sequence that implements the PR), recording each commit's SHA, author, timestamp, and message.

For each changed file, the pipeline retrieves its path, change status, previous path when renamed, numbers of added and deleted lines, and textual patch. In addition, it downloads the complete file contents at the stored base and head revisions. %

The retrieved metadata are stored in a \texttt{data.json} file together with data about the associated issue, while changed files are downloaded locally. %
Materialization completed for all 13,192 validated issues, with no records discarded at this stage.

\subsection{Literal Verification}
We finally retain only the records for which the complete user story \emph{occurs verbatim} in the original issue text. This stage reconstructs the entire user story using the LLM-extracted components and a predefined set of Connextra connectors. The reconstructed story is then matched against the original issue text, considering both complete stories (role, feature, and benefit) and partial stories (role and feature only).

This verification prevents text manipulation accidentally introduced by the LLM %
and ensures that the extracted user story matches the original issue text. Literal Verification thus guarantees by construction that every record
contains a verbatim Connextra story written by the developers; a manual
inspection of 150 randomly selected records confirmed that the stage behaves
as intended.
This final check retains \numStories records (59.01\% of the materialized corpus). Both the original issue title and body are preserved in the output dataset.

\begin{figure*}[ht]
    \centering

    \begin{subfigure}[b]{0.24\linewidth}
        \centering
        \includegraphics[width=\linewidth]{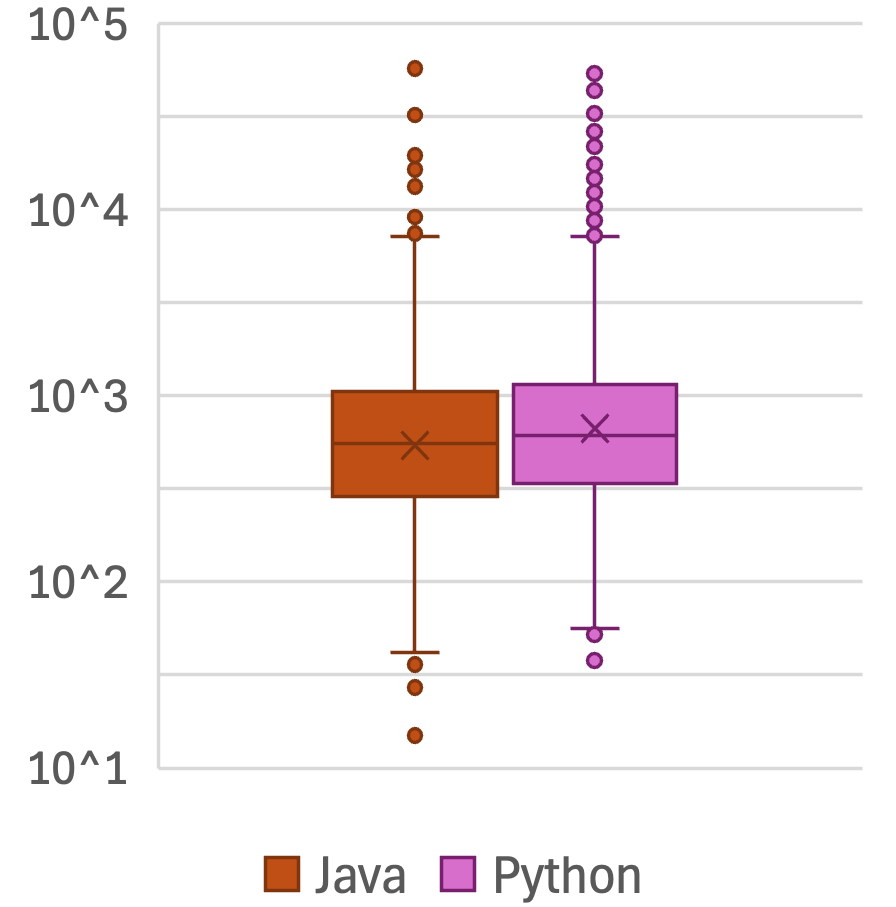}
        \caption{Issue length}
        \label{fig:Issue length}
    \end{subfigure}
    \hfill
    \begin{subfigure}[b]{0.24\linewidth}
        \centering
        \includegraphics[width=\linewidth]{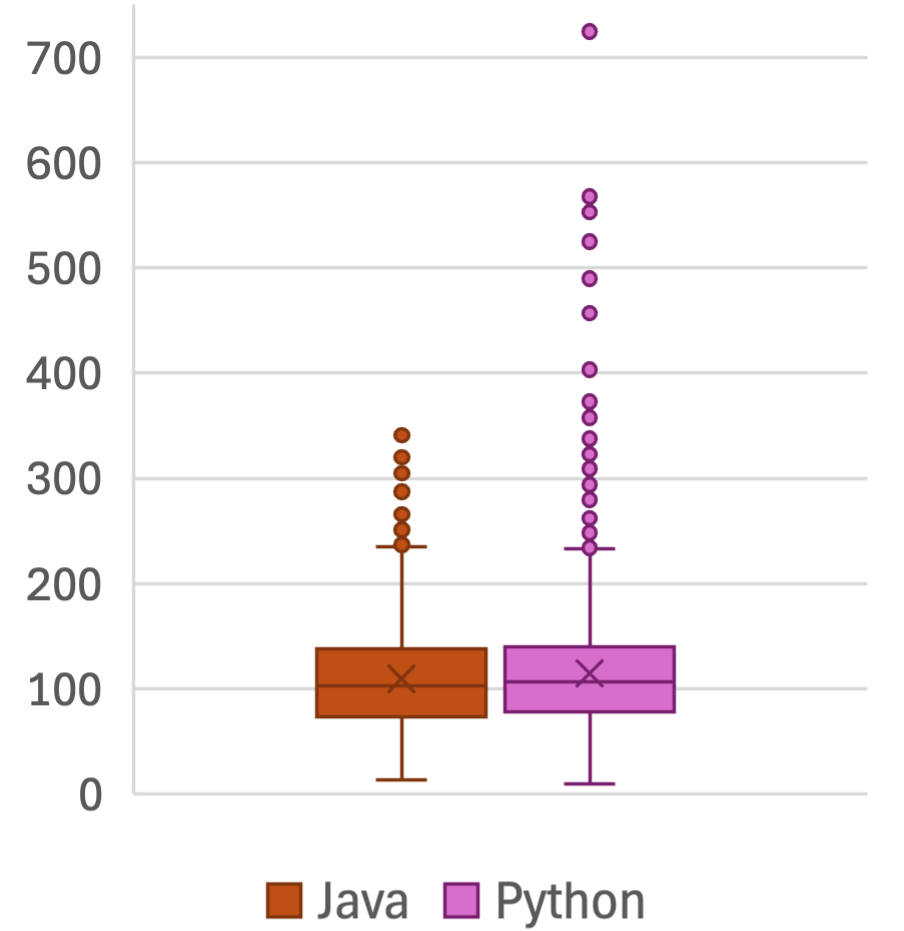}
        \caption{User story length}
        \label{fig:user story length}
    \end{subfigure}
    \hfill
    \begin{subfigure}[b]{0.24\linewidth}
        \centering
        \includegraphics[width=\linewidth]{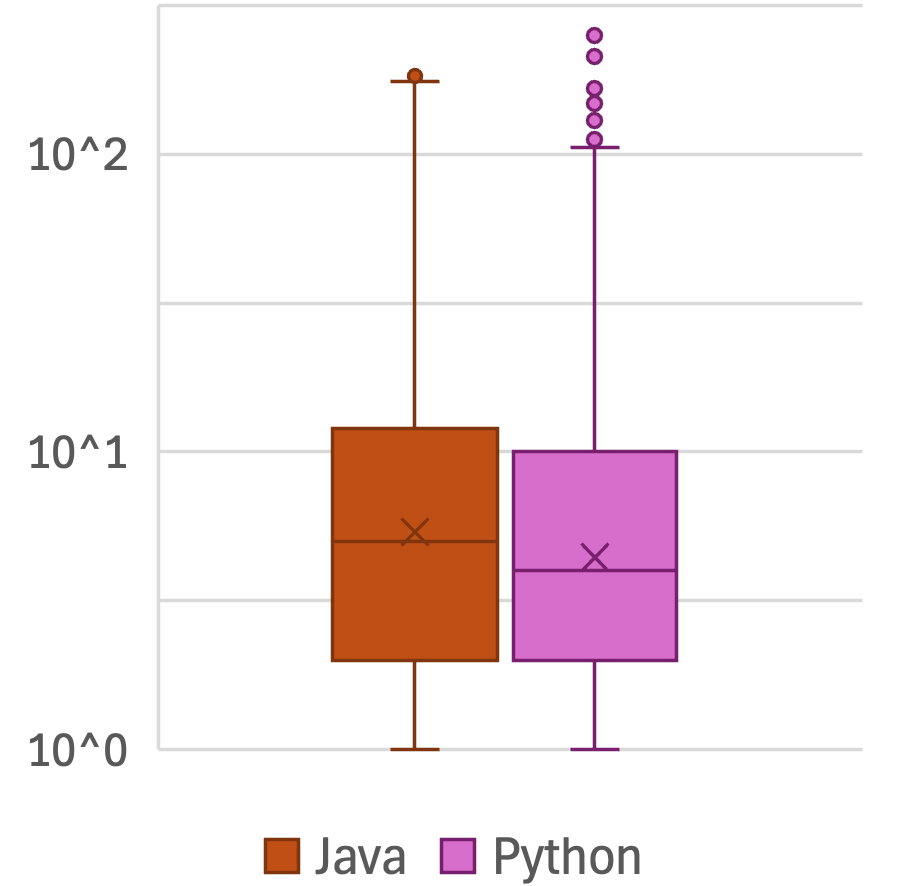}
        \caption{User story commits}
        \label{fig:user story commits}
    \end{subfigure}
    \hfill
    \begin{subfigure}[b]{0.24\linewidth}
    \centering
    \includegraphics[width=\linewidth]{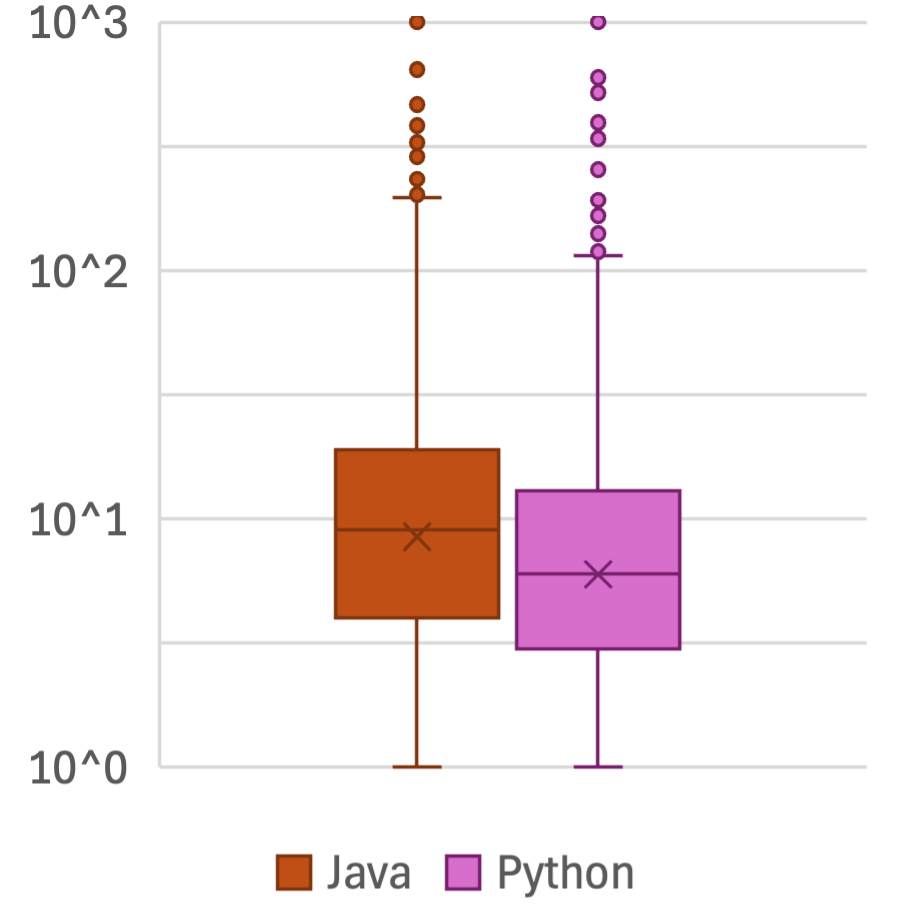}
    \caption{Changed files per user story}
    \label{fig:Changed file per user story}
\end{subfigure}
    \caption{Overview of user story length, issue length, commits and changes}
    \label{fig:user story overview}
    \vspace{-4mm}
\end{figure*}

\section{Data Organization and Tool Support}
The dataset is organized hierarchically by language, repository, and issue. Each issue folder contains the issue and associated PR data and metadata. The GitHub URLs and immutable commit SHAs identify the source artifacts.

The \texttt{original\_files/} and \texttt{modified\_files/} folders store file versions before and after the associated PR changes. %
This organization permits both requirement-to-code and code-to-requirement traversal without re-querying GitHub.
We also provide Python scripts to automate the selection process and repeat the analysis with different parameters (e.g., temporal constraints). A Web application allows users to configure and launch the analysis, monitor its progress, and visualize the results, including through plots.

The code and usage instructions are available at \url{\codeurl}. %

\section{Dataset Characterization,  Limitations, and Uses} \label{sec:quality}

The final \dataset release contains 7,785 records from
982 GitHub repositories. %
Data quality is supported by two complementary checks: each issue must be explicitly closed by a merged PR, %
and the extracted user story must occur verbatim in the original issue text. %

The dataset comprises 4,663 Python records (59.9\%) and
3,122 Java records (40.1\%). Figure~\ref{fig:user story overview} characterizes
the textual and implementation-related properties of
the dataset. User stories exhibit similar lengths across
languages, with median values of 103 characters for Java
and 107 for Python, while the corresponding issues are
considerably longer (544 and 611 characters, respectively).
Regarding implementation size, Java records involve
slightly more commits (median 5 versus 4) and modified
files (median 9 versus 6) than Python records.

\begin{figure}[h]
    \centering
    \includegraphics[width=0.9\linewidth]{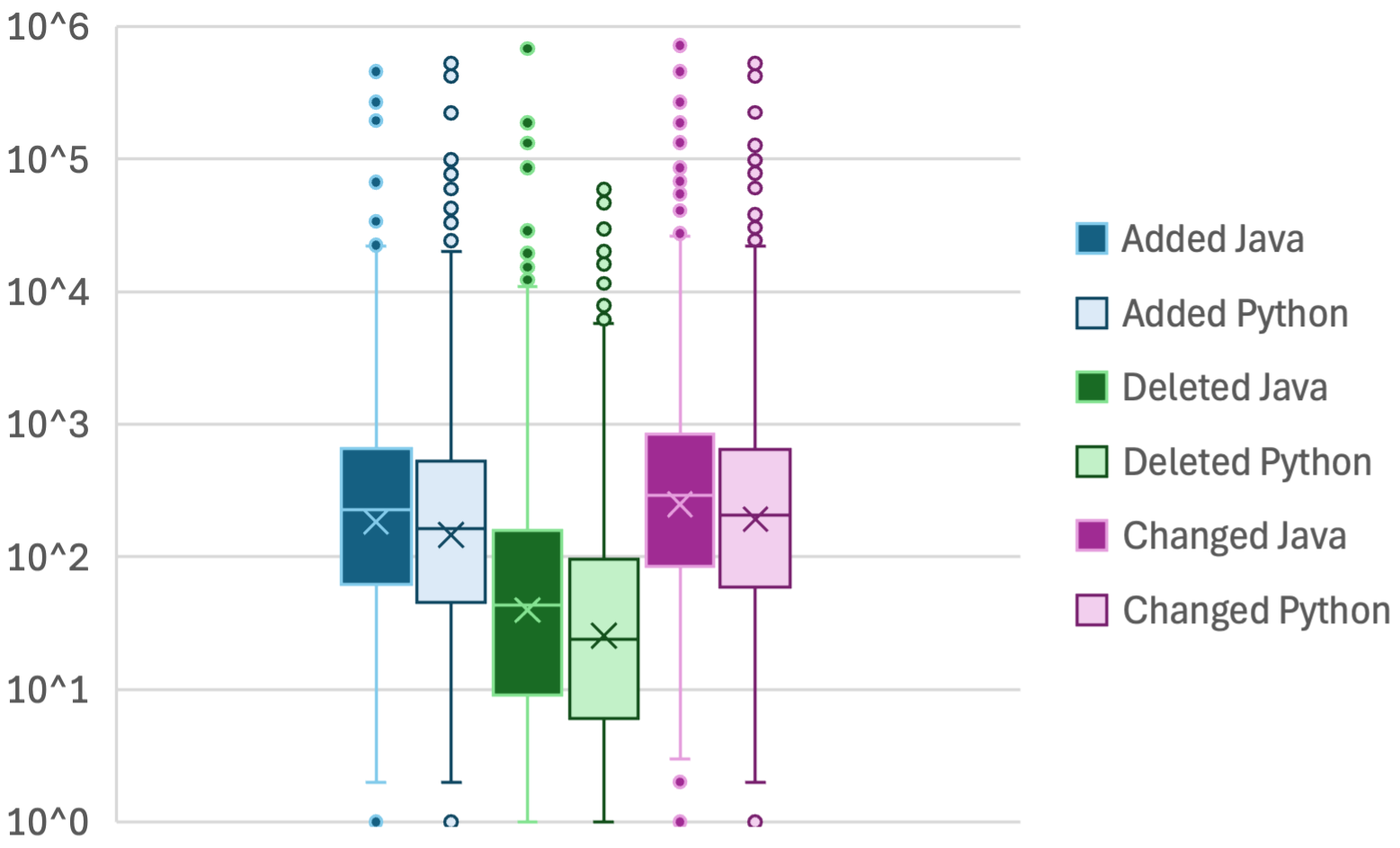}
    \caption{Lines changed per user story}
    \label{fig:Lines changed per user story}
    \vspace{-3mm}
\end{figure}

Figure~\ref{fig:Lines changed per user story} further characterizes the extent of code
modifications. Java records exhibit higher median values
for additions, %
deletions, %
and changes. %
Both figures reveal
substantial variability and numerous outliers, indicating
that the dataset covers implementations of widely
different sizes.

\emph{Limitations.} The dataset is restricted to public Java and Python repositories and to user stories following the supported Connextra patterns.  %
Literal verification ensures textual correspondence, but does not establish requirement completeness or implementation correctness. Moreover, associated PRs may contain changes unrelated
to the user story. %

\emph{Uses.} \dataset supports several research tasks, including the following cases.
\textbf{Requirement recovery:} generate user stories from code and compare them with the original requirements.
\textbf{Traceability:} train or evaluate requirement--PR--code link creation, evolution, and recovery techniques.
\textbf{Effort and size prediction:} relate user-story features to code churn and development time.
\textbf{Automated Code Generation:} compare generated code with the changes associated with a user story.
 \textbf{Project evolution:} 
 analyze the commit-level evolution of code changes associated with user stories.

    \vspace{-2mm}

\bibliographystyle{IEEEtran}
\bibliography{references}

\end{document}